\documentclass[10pt,twocolumn]{article}
\usepackage[margin=0.85in,columnsep=0.3in]{geometry}
\usepackage{amsmath,amssymb}
\usepackage{graphicx}
\usepackage{booktabs}
\usepackage[table]{xcolor}
\usepackage{caption}
\usepackage{hyperref}
\usepackage{titlesec}
\usepackage{enumitem}
\usepackage{authblk}
\usepackage{abstract}
\usepackage{fancyhdr}
\usepackage{float}

\definecolor{corvicpurple}{HTML}{7C3AED}
\definecolor{corvicpurpledark}{HTML}{5B21B6}
\definecolor{corvicsoft}{HTML}{F3EDFC}

\hypersetup{colorlinks=true, linkcolor=corvicpurpledark, citecolor=corvicpurpledark, urlcolor=corvicpurpledark}

\titleformat{\section}{\normalfont\large\bfseries}{\thesection.}{0.5em}{}
\titleformat{\subsection}{\normalfont\bfseries\itshape}{\thesubsection}{0.5em}{}
\titlespacing*{\section}{0pt}{1.4ex plus 1ex minus .2ex}{0.9ex plus .2ex}
\titlespacing*{\subsection}{0pt}{1.1ex plus .8ex minus .2ex}{0.6ex plus .2ex}

\newcommand{\corvic}{\textbf{Corvic AI}}

\title{\Large\bfseries Graph Memory for LLM Agents: At What Cost?\\[2pt]
\large A Comparative Evaluation of Query, Ingest, and Update Performance Across Graph Database Engines}
\author[1]{Donald Nguyen}
\author[1]{Gurbinder Gill}
\author[1]{Hadi Ahmadi}
\author[2]{Christopher J. Rossbach}
\affil[1]{Corvic AI Research \textit{\{ddn, gill, hadi\}@corvic.ai}}
\affil[2]{UT Austin \textit{rossbach@cs.utexas.edu}}
\date{}

\begin{document}
\twocolumn[
\begin{@twocolumnfalse}
\maketitle
\begin{abstract}
\noindent Graph databases are frequently positioned as categorically necessary for connected-data workloads, yet the systems dimension along which they actually differ---query planning, indexing, and data-readiness cost---is rarely isolated from vendor framing. We construct a synthetic, biomedical-shaped property graph (1.02~million nodes, 5.34~million total node and edge rows) and a twenty-query workload spanning neighborhood lookups, bounded paths, set intersections, anti-joins, grouped aggregation, top-$k$ ranking, temporal filters, full scans, and relational joins. We benchmark \corvic{}---a purpose-built columnar query engine underlying Corvic's ontology management layer (``memories'')---against seven purpose-built or graph-extension database systems (LoraDB, Ladybug, DuckPGQ, Memgraph, Neo4j, HugeGraph, and FalkorDB) at three graph scales spanning three orders of magnitude. We report query latency geomeans, bulk-ingest throughput, point-update latency, and answer correctness for each system, and we derive a simple total-cost-of-ownership model that expresses the ingest/query trade-off as a function of query volume. Our central finding is that no system in this sample is categorically fastest: a native graph engine (Ladybug) outperforms Corvic AI on narrow, bounded-neighborhood shapes, while Corvic AI is faster on shapes that scan or join a large fraction of the graph, and a system implementing graph query syntax via SQL/PGQ (DuckPGQ) is measurably slower purely due to query-plan choice. The dominant cost differential in our data is not query latency but the cost of making data queryable at all: bulk-ingest throughput varies by three orders of magnitude across engines (5.0k--4.3M~rows/s), a gap that a simple crossover-point calculation shows dominates total cost for any workload with fewer than roughly $10^{5}$ queries per data refresh.

\end{abstract}
\vspace{1em}
\end{@twocolumnfalse}
]

\section{Introduction}

When Google introduced its Knowledge Graph in 2012, the framing was ``things, not strings''---a claim that search, and by extension much of applied data systems, should reason over typed entities and relationships rather than over token sequences \cite{google2012}. The claim reads as almost contrarian today: contemporary interfaces to data are overwhelmingly conversational, and the dominant interface pattern---the prompt---is a string.

A prompt can name a company, but on its own it cannot assert that the company is a customer, that a contract governs the account, that an incident affected one of its deployments, or that the entity issuing the query is authorized to see the answer. Nor can it represent how such facts change over time. These are structured claims---entities and the relationships between them---and as language-model-driven systems move from answering questions about text to acting inside operational business processes, they require a substrate that represents this structure explicitly rather than inferring it transiently from a context window~\cite{lewis2020rag,pan2024unifyingllmkg,edge2024graphrag}.

An \emph{ontology} is commonly defined as a formal, explicit specification of a shared conceptualization of a domain: a set of concepts, properties, and relationships that a community of use agrees represent the entities and connections that matter within that domain~\cite{gruber1993ontology}. At Corvic, we build and maintain such ontologies under the name \emph{memories}; the query engine underlying this layer is \corvic{}, a purpose-built columnar analytical query engine. This paper reports an empirical comparison of that engine's query, ingest, and update performance against a sample of purpose-built graph database systems, undertaken to understand the engineering trade-offs relevant to that choice---not to declare a universally superior system.

\subsection{Contributions and scope}

We make the scope of this study deliberately narrow. We do not claim comprehensiveness over the graph database market, and we explicitly reject the framing of ``best database'' benchmarks common in vendor literature. Instead we ask: \emph{what dimensions of variation among graph-capable systems are load-bearing, and which are incidental?} Concretely, we contribute (i) a reproducible workload of twenty query shapes designed to separate query-planning effects from storage-engine effects~\cite{selinger1979}, (ii) latency, throughput, and correctness measurements for eight systems at three graph scales, (iii) a decomposition of one query family (one-hop neighbor lookup) into query-plan artifacts that explain an observed 2.3$\times$ latency gap between two SQL-based engines applying different query-plan strategies to the same query shape, and (iv) a total-cost-of-ownership model that makes the ingest/query trade-off explicit as a function of query volume, together with a worked numerical example.

\section{Related Work and Positioning}

Graph database benchmarking has a substantial existing literature, most visibly the LDBC Social Network Benchmark \cite{ldbc2020} and Graphalytics \cite{graphalytics2016} suites, which standardize workloads across native graph engines (Neo4j, TigerGraph, JanusGraph) and graph-processing frameworks~\cite{malewicz2010pregel}. These benchmarks are valuable but structurally oriented toward comparing graph systems \emph{to each other}~\cite{besta2023demystifying,sakr2021futurebiggraphs}; they do not typically include general-purpose columnar OLAP engines as a baseline, and so cannot speak to whether ``graph-native'' storage is a necessary condition for graph-shaped query performance~\cite{angles2008survey}. A parallel literature on relational encodings of graph data~\cite{codd1970}---property graphs over relational schemas, SQL/PGQ (ISO/IEC~39075:2024) \cite{sqlpgq2024}, and extensions such as DuckPGQ \cite{duckpgq2024}---asks a closer question: whether a graph query language can be compiled to a relational plan without loss of performance. Our design borrows from both traditions: we adopt a workload-shape taxonomy in the spirit of LDBC-style diversity (neighborhood, path, aggregation, and join queries) while including both graph-native systems and a relational engine (Corvic AI) and a relational-with-graph-syntax system (DuckPGQ) side by side, which lets us separate \emph{syntax} from \emph{execution plan} as an experimental variable~\cite{francis2018cypher,rodriguez2015gremlin} (Section~\ref{sec:duckpgq}).

Our aim in what follows is not to identify a single best system among those tested, but to characterize how these systems tend to vary along the dimensions that determine practical suitability: query-plan quality, ingest throughput, and update cost. Section~\ref{sec:duckpgq} in particular treats variation in query-plan choice as the object of study, holding query language and workload shape fixed, so that any difference in outcome is attributable to planning rather than to interface; Section~\ref{sec:ingest} does the same for the ingest/update trade-off, holding the workload fixed while varying the engine.

\section{Experimental Setup}

\subsection{Dataset}

We generated a synthetic property graph with a biomedical-knowledge-graph shape: typed node collections analogous to \texttt{gene}, \texttt{protein}, \texttt{disease}, and \texttt{compound} entities, connected by typed, directed and undirected edge collections analogous to interaction, association, and regulation relationships. This entity and relation taxonomy was designed with reference to the schema of the STaRK-PRIME semi-structured retrieval benchmark~\cite{wu2024stark}, which is itself built on the PrimeKG precision-medicine knowledge graph~\cite{chandak2023primekg}; the real STaRK-PRIME dataset is publicly available for download at \url{https://stark.stanford.edu/dataset_prime.html}. We emphasize that our benchmark graph is synthetically generated at the scales described below and is not a sample, subset, or derivative of PrimeKG or STaRK-PRIME itself (see Limitations, Section~\ref{sec:limitations}). The full graph comprises 1.02~million nodes and 5.34~million total node and edge rows. To study how each system's performance changes with data volume, we additionally materialize two down-sampled scales of the same schema and generative process, at 1{,}000 and 10{,}000 nodes, giving three scales spanning approximately three orders of magnitude.

\subsection{Query workload}

We designed twenty query shapes intended to be jointly exhaustive over the query patterns we observe in production ontology workloads: one-hop and multi-hop neighborhood lookups, bounded-length path queries (\texttt{chain\_paths}), set intersections, anti-joins (existence-negated lookups), grouped counting aggregations, top-$k$ ranked retrieval (\texttt{top\_k\_ordered})~\cite{fagin2001topk}, temporal range filters, unfiltered scans, and star/chain relational joins including a large equi-join across two high-cardinality tables (\texttt{hash\_join}) and a self-referential join (\texttt{self\_join}). Each shape is implemented natively in each engine's own query language (SQL, SQL/PGQ, Cypher, Gremlin, or engine-specific DSL as applicable) rather than through a single translated query string, so that each engine is exercised through its own idiomatic access path.

\subsection{Systems under test}

We evaluate eight systems, chosen to span a spectrum from general-purpose columnar OLAP~\cite{stonebraker2005cstore,boncz2005x100} to dedicated native graph storage, and spanning a wide range of project maturity: \corvic{} (our subject system, a purpose-built columnar query engine), \textbf{DuckPGQ} (a SQL/PGQ extension for relational query engines), \textbf{LoraDB}, \textbf{Ladybug}, \textbf{Memgraph}, \textbf{Neo4j}, \textbf{HugeGraph}, and \textbf{FalkorDB}. A ninth system, PuppyGraph, was evaluated but is excluded from all reported figures because it could not complete the query workload at the tested scales under our harness; see Appendix~\ref{app:a}. Ladybug supports two data-attachment configurations (\texttt{attach}, which references source files in place, and \texttt{copy}, which materializes an internal copy); we report only the better-performing \texttt{copy} configuration in the main text to avoid double-counting one system, and disclose \texttt{attach} figures in Appendix~\ref{app:a}.

\subsection{Measured quantities}

For each system, at each of the three scales, we record: (i) per-query latency for all twenty query shapes, from which we derive a geometric-mean summary statistic (Eq.~\ref{eq:geomean}); (ii) whether each query's result was verified correct against a reference implementation; (iii) end-to-end bulk-ingest throughput when loading the full node and edge set from cold storage; (iv) single-record point-update latency; and (v) batch-update throughput for a fixed-size batch of mutations applied after initial load. All reported figures are computed over repeated runs on identical hardware under a fixed harness; ms figures throughout are the per-engine geometric mean over the twenty query shapes at the stated scale unless otherwise noted.

\section{Formal Definitions}

We define the summary statistics used throughout the paper explicitly, since a benchmark's conclusions are frequently sensitive to the choice of aggregation.

\subsection{Query latency geomean}

For engine $i$ with per-shape latencies $t_{i,1},\ldots,t_{i,n}$ over the $n=20$ query shapes at a given scale, we report the geometric mean rather than the arithmetic mean, since query latencies are heavy-tailed (our slowest shape is roughly 400$\times$ our fastest for the same engine) and the arithmetic mean would be dominated by outlier shapes unrepresentative of typical workload composition~\cite{fleming1986geomean}:
\begin{equation}
\label{eq:geomean}
GM_i = \left( \prod_{k=1}^{n} t_{i,k} \right)^{1/n}
\end{equation}
Where a query cell did not complete or returned an incorrect result, that cell is excluded from the product and $n$ is reduced accordingly for that engine; we mark all such rows with the correct/total ratio (Table~\ref{tab:main}, Appendix~\ref{app:a}).

\subsection{Ingest throughput}

Ingest throughput is total rows loaded (nodes and edges combined) divided by wall-clock ingest time~\cite{stonebraker2005cstore}:
\begin{equation}
\label{eq:ingest}
R_i = \frac{N_{\text{rows}}}{T_{\text{ingest},i}}
\end{equation}

\subsection{Relative slowdown}

To compare engines at a fixed scale independent of absolute units, we define the slowdown of engine $i$ relative to Corvic AI at scale $s$ as the ratio of geomeans:
\begin{equation}
\label{eq:slowdown}
S_{i,s} = \frac{t_{i,s}}{t_{\text{corvic},s}}
\end{equation}
By construction $S_{\text{corvic},s}=1$ for all $s$; Figure~\ref{fig:heatmap} reports this quantity as a heatmap over all engines and scales.

\subsection{Correctness rate}

We define per-engine correctness at scale $s$ as the fraction of the twenty query shapes returning a result verified against a reference implementation:
\begin{equation}
\label{eq:accuracy}
A_i = \frac{c_i}{20}\times 100\%
\end{equation}

\subsection{Total cost of ownership as a function of query volume}

A benchmark that reports only query latency implicitly assumes the data is already loaded and never changes---an assumption that does not hold for an actively maintained ontology~\cite{mcsherry2015cost}. We instead model the total wall-clock cost of using engine $i$ for a workload of $q$ queries issued against a freshly (re-)ingested dataset as the sum of one-time ingest cost and per-query cost at mean latency $\bar t_i$:
\begin{equation}
\label{eq:tco}
TCO_i(q) = T_{\text{ingest},i} + q\cdot \bar t_i
\end{equation}
Given two engines $A$ and $B$ with $T_{\text{ingest},A} < T_{\text{ingest},B}$ but $\bar t_A > \bar t_B$ (the faster-to-load engine is slower per query, or vice versa), the two cost curves cross at a query volume $q^{*}$ solving $TCO_A(q^{*}) = TCO_B(q^{*})$:
\begin{equation}
\label{eq:crossover}
q^{*} = \frac{T_{\text{ingest},B} - T_{\text{ingest},A}}{\bar t_A - \bar t_B}
\end{equation}
Below $q^{*}$, the engine with lower ingest cost has lower total cost regardless of its per-query latency; above $q^{*}$, the reverse holds. Section~\ref{sec:ingest} instantiates this model numerically for Corvic AI versus Memgraph.

\section{Results}

\subsection{Query latency at full scale}

\begin{figure}[H]
\centering
\includegraphics[width=\linewidth]{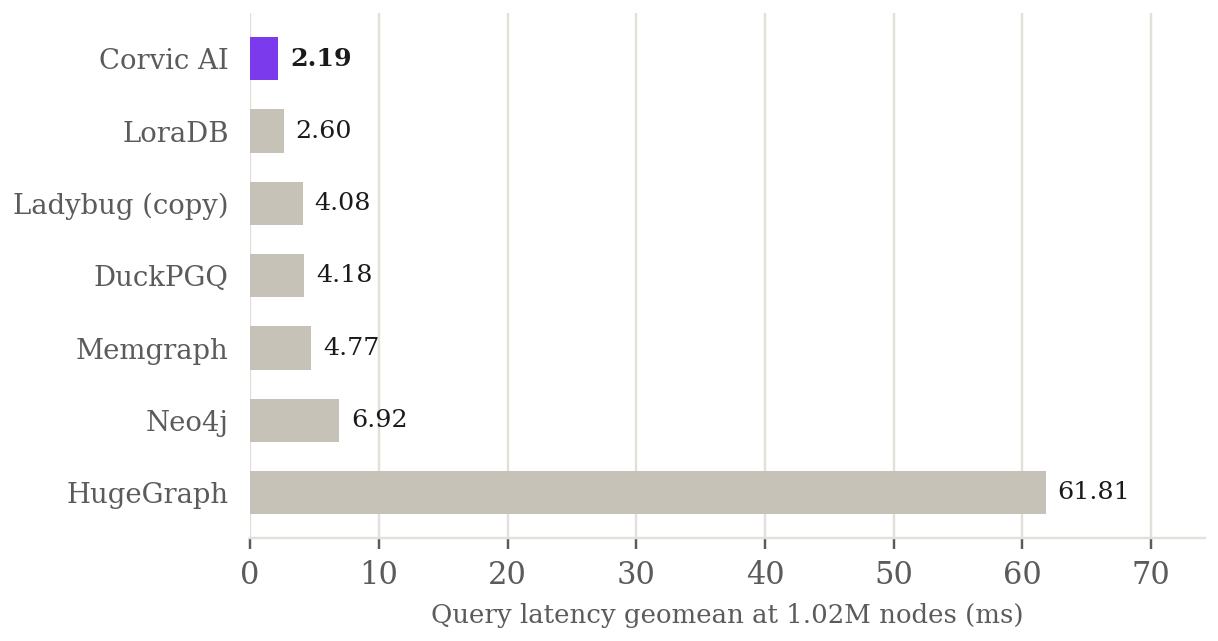}
\caption{Query-latency geomean at 1.02M nodes (Eq.~\ref{eq:geomean}), all eight engines, lower is better. Corvic AI shown in purple.}
\label{fig:fig1}
\end{figure}

Table~\ref{tab:main} reports all four measured quantities for all eight engines at the 1.02M-node scale. Figure~\ref{fig:fig1} visualizes the query-latency geomean column. Corvic AI's geomean (2.19~ms) is the second-lowest of the eight systems tested; FalkorDB is lower (1.81~ms) but completed only 18 of 20 shapes correctly (see Eq.~\ref{eq:accuracy} and Appendix~\ref{app:a}), so the two figures are not directly comparable on a correctness-adjusted basis. Among engines completing all twenty shapes correctly, Corvic AI is fastest, followed by LoraDB (2.60~ms, 18/20 correct), Ladybug (3.97~ms, 20/20), DuckPGQ (4.18~ms, 20/20), Memgraph (4.77~ms, 20/20), Neo4j (6.92~ms, 20/20), and HugeGraph (61.8~ms, 19/20), the last being 28.2$\times$ slower than Corvic AI at this scale.

\subsection{Scaling behavior across three orders of magnitude}

\begin{figure}[H]
\centering
\includegraphics[width=\linewidth]{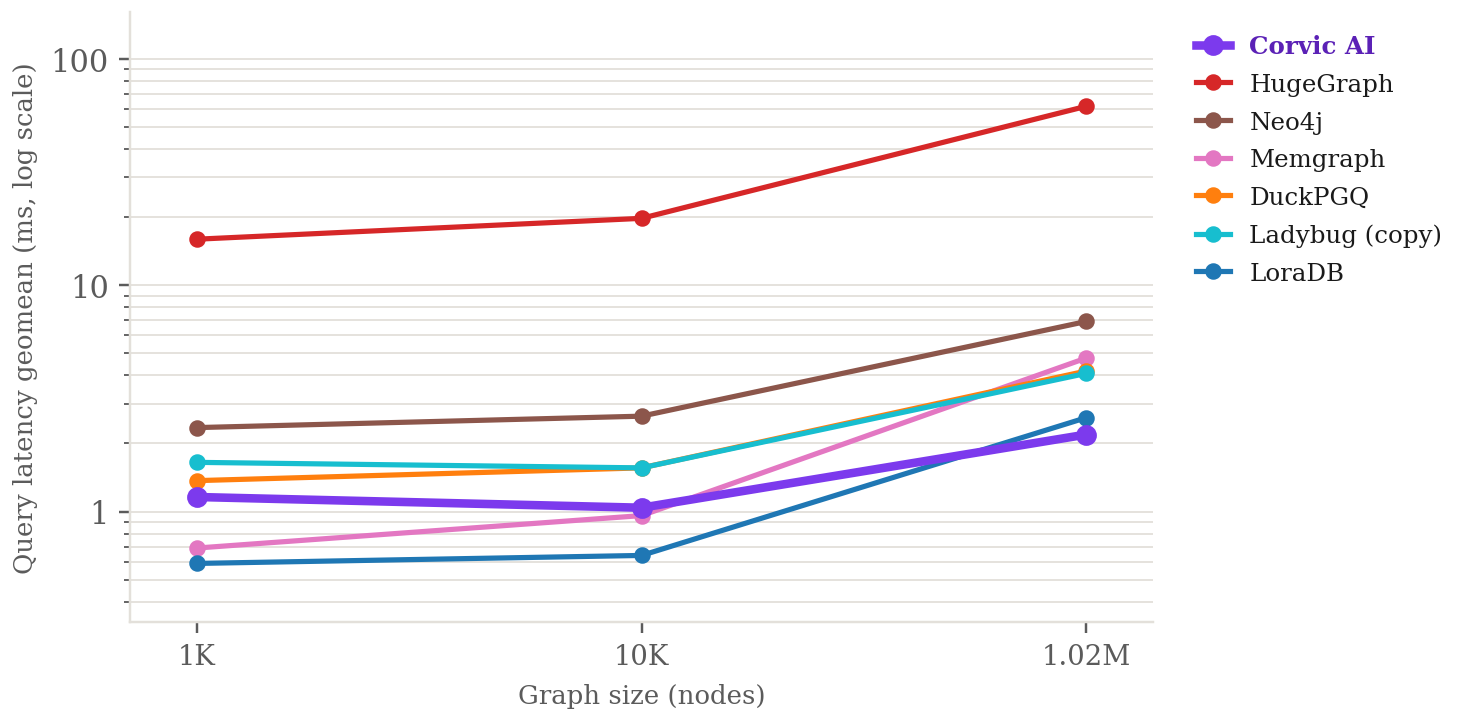}
\caption{Query-latency geomean versus graph scale (y log, x categorical), all eight engines. Corvic AI drawn with a heavier line.}
\label{fig:fig2}
\end{figure}

Figure~\ref{fig:fig2} plots the same geomean statistic at all three tested scales (1K, 10K, and 1.02M nodes) on a log axis. Two qualitative patterns are notable. First, rank order is not stable across scale: LoraDB and Memgraph post the lowest latencies at 1K and 10K nodes (0.59--0.69~ms) but fall behind Corvic AI by the 1.02M-node scale, indicating their query-planning or indexing overhead scales worse with data volume than Corvic AI's columnar scan-based execution~\cite{abadi2008columnstore,boncz2005x100}. Second, HugeGraph's latency grows fastest in absolute terms (15.98~ms~$\to$~61.81~ms across our three scales), consistent with an execution strategy whose cost grows superlinearly with graph size on at least a subset of our twenty shapes.

\subsection{Query-plan sensitivity: DuckPGQ versus Corvic AI}
\label{sec:duckpgq}

DuckPGQ is a SQL/PGQ extension for relational query engines; comparing it against Corvic AI's handwritten SQL lets us examine how much of the latency difference between a graph-query-syntax interface and a hand-written relational query is attributable to query-plan choice rather than to query language design. We find the two engines are effectively tied on our relational-join query shapes (1.26~ms geomean for Corvic AI's handwritten SQL versus 1.20~ms for DuckPGQ), but diverge sharply on graph-family shapes (2.52~ms versus 5.70~ms, a 2.26$\times$ gap)---a difference driven by \texttt{GRAPH\_TABLE} plan choices rather than by SQL/PGQ syntax per se.

The query plan for a representative one-hop \texttt{neighbors} shape illustrates the mechanism. Corvic AI's handwritten SQL scans a filtered edge table directly. DuckPGQ's \texttt{GRAPH\_TABLE} plan instead performs an explicit three-way join across the source-vertex, edge, and destination-vertex tables, scanning all 330{,}000 rows of the destination (\texttt{protein}) table to confirm each destination carries the requested label---even though the edge table's own foreign key already identifies the destination row without requiring that scan. What is logically a single filtered scan becomes, in the executed plan, two hash joins. We emphasize that this is a property of the specific plan chosen by the current DuckPGQ optimizer for this shape, not an inherent property of SQL/PGQ as a language: a cost-based optimizer aware of the foreign-key constraint could, in principle, produce the same plan Corvic AI's handwritten SQL executes directly~\cite{selinger1979}.

\subsection{Shape-dependent wins for a native graph engine}

Ladybug, a purpose-built native graph engine, outperforms Corvic AI specifically on shapes bounded to a small local neighborhood: 1.76~ms versus 2.62~ms on \texttt{chain\_paths}, a bounded-length path query. This is consistent with native adjacency-list storage offering a genuine advantage when a query only needs to touch a small, localized subgraph and can avoid the columnar engine's per-scan overhead entirely~\cite{besta2023demystifying}. The same engine, however, loses substantially on shapes that touch a large fraction of the graph: 13.6~ms versus 1.92~ms on \texttt{self\_join} (7.1$\times$ slower), 36.0~ms versus 7.8~ms on \texttt{hash\_join} (4.6$\times$ slower), and 49.0~ms versus 4.5~ms on \texttt{top\_k\_ordered} (10.9$\times$ slower). We read this pattern as the clearest single piece of evidence in our data that ``graph-native'' storage is a targeted optimization for a specific access pattern (bounded local traversal) rather than a general performance advantage over columnar scan-and-join execution~\cite{sakr2021futurebiggraphs}.

\begin{quotation}
\noindent\itshape\color{corvicpurpledark!80!black}
No single engine in our sample dominates on every query shape we tested; the shape of the query, not the label ``graph database,'' is the better predictor of which engine will be fastest.
\end{quotation}

\subsection{Ingest, update cost, and the total-cost crossover}
\label{sec:ingest}

\begin{figure}[H]
\centering
\includegraphics[width=\linewidth]{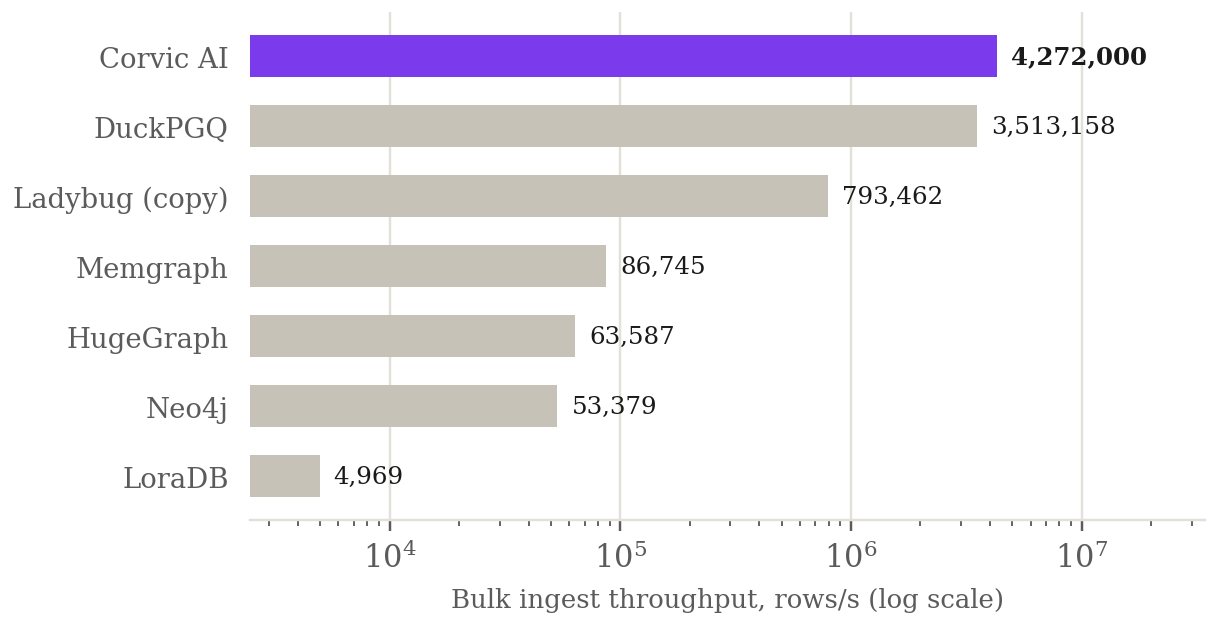}
\caption{Bulk-ingest throughput at 1.02M nodes, rows/s, log scale, higher is better.}
\label{fig:fig4}
\end{figure}

\begin{figure}[H]
\centering
\includegraphics[width=\linewidth]{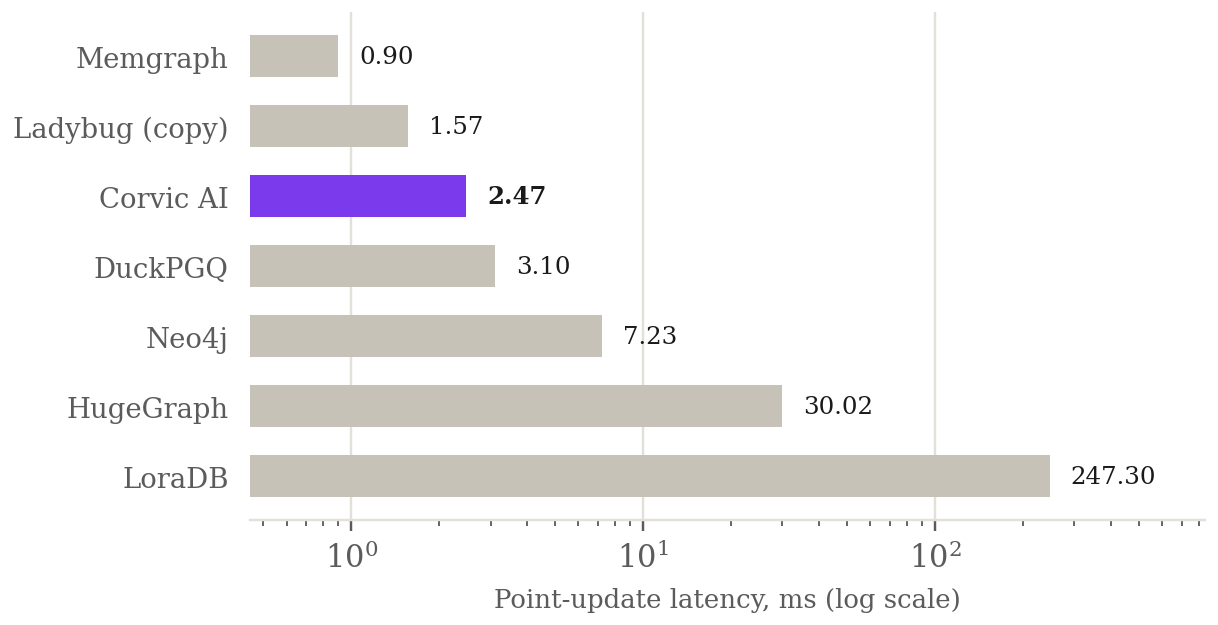}
\caption{Point-update latency at 1.02M nodes, ms, log scale, lower is better. The only workload in this study where Corvic AI does not lead.}
\label{fig:fig5}
\end{figure}

Figure~\ref{fig:fig4} shows bulk-ingest throughput at the full 1.02M-node scale. The spread is the largest effect size we measure anywhere in this study: 4.3M~rows/s for Corvic AI down to 5.0k~rows/s for LoraDB, a factor of roughly 860$\times$ between the fastest and slowest engine we benchmarked, and clearly the dominant source of variation across our whole result set---larger than any query-latency gap in Table~\ref{tab:main}. Figure~\ref{fig:fig5} shows point-update latency, the one workload in this study on which Corvic AI is not the fastest engine: Memgraph (0.90~ms) and FalkorDB (0.85~ms) both edit a single existing record faster than Corvic AI (2.47~ms), consistent with these systems' in-memory, mutation-oriented storage design.

\begin{figure}[H]
\centering
\includegraphics[width=\linewidth]{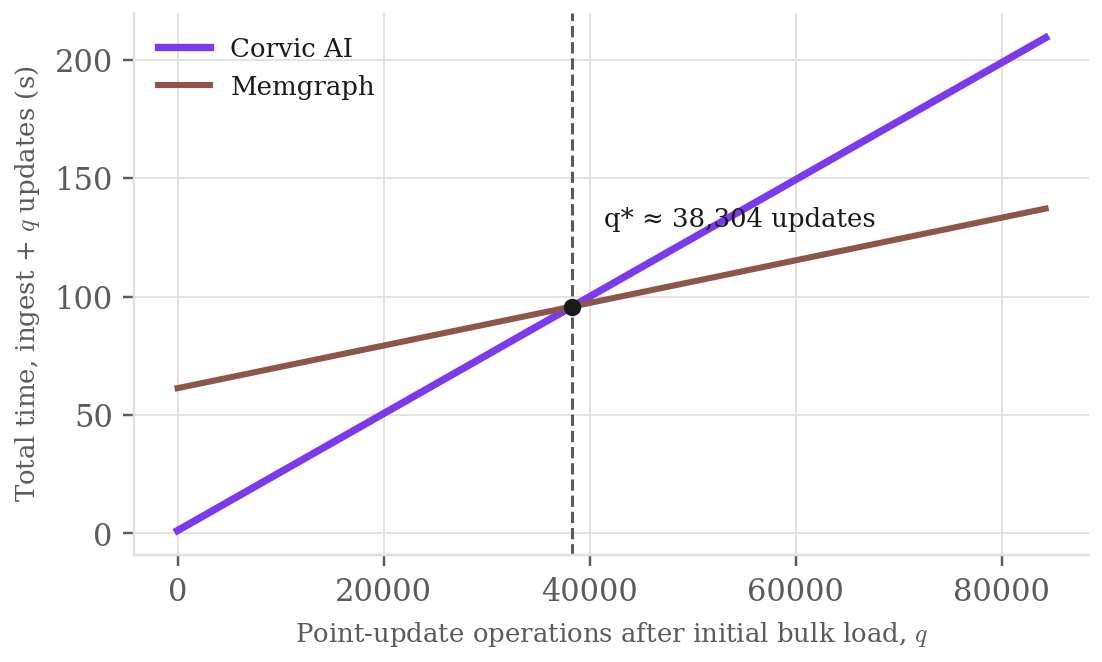}
\caption{Total wall-clock cost (Eq.~\ref{eq:tco}) versus number of point-update operations issued after a fresh bulk load, Corvic AI versus Memgraph. Dashed line marks the crossover $q^{*}$ (Eq.~\ref{eq:crossover}).}
\label{fig:fig6}
\end{figure}

We apply the TCO model of Section~4.5 to Corvic AI versus Memgraph---the engine that beats Corvic AI on point-update latency---treating ``query'' in Eq.~\ref{eq:tco} as one point-update operation. With $N_{\text{rows}}=5.34\text{M}$, $T_{\text{ingest,corvic}}=1.24$~s, $T_{\text{ingest,memgraph}}=61.4$~s, $\bar t_{\text{corvic}}=2.47$~ms, and $\bar t_{\text{memgraph}}=0.90$~ms per update, Eq.~\ref{eq:crossover} gives a crossover at $q^{*}\approx 38{,}304$ point-update operations: below this volume, Corvic AI's much lower ingest cost dominates total wall-clock time even though each individual update is slower; above it, Memgraph's per-update speed eventually wins out. Figure~\ref{fig:fig6} plots both cost curves. For context, our own operational experience with actively maintained ontologies (Section~\ref{sec:discussion}) is that a dataset is re-ingested or substantially re-augmented far more frequently than 38{,}304 point mutations accumulate against a stable schema between refreshes, which is the empirical condition under which this crossover analysis favors the ingest-optimized engine.

\subsection{Relative slowdown across scale}

\begin{figure}[H]
\centering
\includegraphics[width=0.92\linewidth]{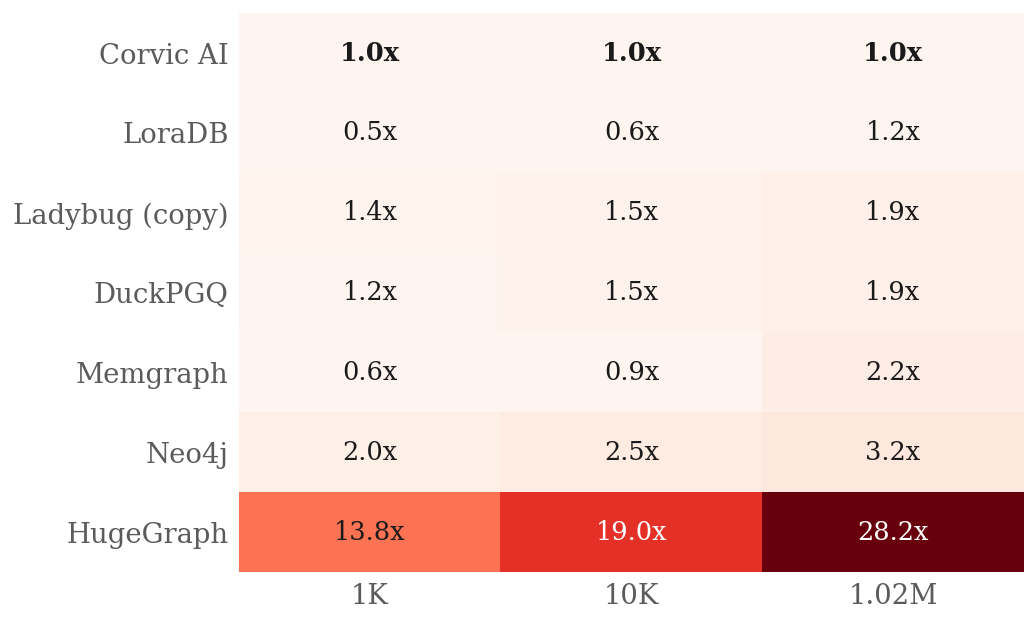}
\caption{Slowdown factor $S_{i,s}$ relative to Corvic AI (Eq.~\ref{eq:slowdown}), by engine and scale. 1.0$\times$ (Corvic AI's own row) is the reference.}
\label{fig:heatmap}
\end{figure}

Figure~\ref{fig:heatmap} reports the slowdown ratio $S_{i,s}$ (Eq.~\ref{eq:slowdown}) for every engine at every scale, normalized to Corvic AI. For most competitor engines the ratio grows with scale rather than shrinking, i.e. the gap widens as the graph grows, which is the regime in which production ontology workloads are more likely to operate.

\section{Discussion}
\label{sec:discussion}

The result we consider most load-bearing for practitioners is not any single latency number in Table~\ref{tab:main} but the shape of the whole picture: query latency, ingest throughput, and update latency trade off against each other in ways that are legible only when reported jointly, and any one of the three, reported alone, supports a different and potentially misleading recommendation. Reported alone, query latency at full scale would suggest Corvic AI and FalkorDB are the strongest choices; reported alone, point-update latency would suggest Memgraph or FalkorDB; reported alone, ingest throughput would strongly favor Corvic AI or DuckPGQ. The engineering decision an organization actually faces---which system to build an ontology layer on---depends on the ratio of query volume to data-change volume for its specific workload, which is precisely the quantity the crossover point $q^{*}$ (Eq.~\ref{eq:crossover}) makes explicit.

For our own use case---ontology management under the ``memories'' system---we find empirically that loading, bulk updating, and augmentation pipelines over an ontology constitute a substantial share of total system load, and that it is uncommon for a given ontology to receive on the order of $10^{5}$ queries (our estimated crossover order of magnitude for millisecond-class queries against a corpus ingested at $10^{4}$~rows/s) without an intervening reload or substantial update. Under that operating regime, ingest and update cost are not a secondary concern to be amortized away; they are frequently the dominant cost. This is, we believe, systematically under-weighted in query-latency-only benchmark reports, including many of our own prior internal evaluations.

The DuckPGQ comparison (Section~\ref{sec:duckpgq}) is, in our view, the most conceptually important result in the paper, because it isolates query-plan quality as a variable independent of storage engine, indexing strategy, or hardware. It demonstrates that adopting graph query syntax (SQL/PGQ, or by extension Cypher~\cite{francis2018cypher} or Gremlin~\cite{rodriguez2015gremlin}) does not by itself guarantee a graph-optimized execution plan; logically equivalent graph and SQL queries produced measurably different plans purely as a function of which query interface was used to express the request, independent of the query language itself. We think this generalizes beyond DuckPGQ: a system's query language is a poor proxy for its query optimizer's quality, and benchmark methodology that varies the query language alongside the storage engine (as most publicly available graph benchmarks do) conflates two variables that this study's design was intended to separate.

\section{Threats to Validity and Limitations}\label{sec:limitations}

\begin{itemize}[leftmargin=1.1em,itemsep=2pt,topsep=2pt]
\item \textbf{Sample size and generality.} Eight engines and one synthetic dataset shape is not a comprehensive survey of the graph database market; our explicit goal (Section~1.1) is to characterize dimensions of variation, not to rank all available systems.
\item \textbf{Synthetic data.} Our graph is generated, not drawn from a production biomedical corpus; although its entity/relation taxonomy was designed with reference to the STaRK-PRIME benchmark~\cite{wu2024stark} and the underlying PrimeKG knowledge graph~\cite{chandak2023primekg}, we do not sample from either dataset, and real-world degree distributions, locality, and update patterns in PrimeKG or other production biomedical graphs may differ from our generative process in ways that shift the relative standing of native-graph versus columnar engines.
\item \textbf{Excluded systems.} FalkorDB and PuppyGraph results are partially or fully excluded (Appendix~\ref{app:a}) because they did not complete our full workload under our harness.
\item \textbf{Ladybug configuration selection.} We report only the faster of Ladybug's two data-attachment modes in the main text; the excluded \texttt{attach} configuration is disclosed in Appendix~\ref{app:a}.
\item \textbf{HugeGraph batch-update measurement.} HugeGraph's batch-update throughput could not be measured to completion under our harness at the full scale and is reported as DNF.
\item \textbf{Single-harness measurement.} All timings were collected under one measurement harness on shared hardware; we did not vary hardware configuration, concurrency level, or client-driver overhead~\cite{mcsherry2015cost}.
\item \textbf{Author affiliation.} Corvic AI is both the subject system and the authors' own product; we have tried to offset this conflict of interest through full disclosure of the one workload where our system does not lead and by publishing the underlying comparison table (Appendix~\ref{app:a}) rather than only derived charts.
\end{itemize}

\section{Conclusion}

On the twenty-shape workload evaluated here, a native graph engine was not categorically faster than a columnar OLAP engine at full scale: Corvic AI led on shapes touching a large fraction of the graph and on both ingest and batch-update throughput, while a native graph engine (Ladybug) led on narrowly bounded neighborhood queries and two other systems (Memgraph, FalkorDB) led on single-record point updates. The largest effect size in our data by a wide margin is not query latency but ingest throughput, which varied by roughly three orders of magnitude across the engines tested~\cite{stonebraker2005cstore} and, per our total-cost-of-ownership model (Eq.~\ref{eq:tco}--\ref{eq:crossover}), dominates total system cost for any workload issuing fewer than approximately $10^{5}$ queries between data refreshes---the regime we believe describes most actively maintained ontologies in practice. We release the comparison table underlying every figure in this paper (Appendix~\ref{app:a}) and encourage readers to apply Eq.~\ref{eq:crossover} to their own query-volume and refresh-cadence estimates rather than to adopt our ranking directly.

\onecolumn
\appendix
\section{Appendix --- Full Comparison Table}
\label{app:a}

Table~\ref{tab:main} reports every quantity referenced in the main text for all eight engines evaluated, at the full 1.02M-node / 5.34M-row scale, plus the 1K- and 10K-node query geomeans used in Figure~\ref{fig:fig2}. Figures are as measured under our harness; DNF denotes a measurement that did not complete.

\begin{table}[H]
\centering
\small
\caption{Full per-engine results. Geomean columns per Eq.~\ref{eq:geomean}; correctness per Eq.~\ref{eq:accuracy}; ingest per Eq.~\ref{eq:ingest}. Corvic AI row highlighted.}
\label{tab:main}
\begin{tabular}{lrrrrrrr}
\toprule
\textbf{Engine} & \textbf{GM@1K} & \textbf{GM@10K} & \textbf{GM@1.02M} & \textbf{Correct} & \textbf{Ingest} & \textbf{Point} & \textbf{Batch} \\
 & \textbf{(ms)} & \textbf{(ms)} & \textbf{(ms)} & \textbf{@1.02M} & \textbf{(rows/s)} & \textbf{upd. (ms)} & \textbf{(rows/s)} \\
\midrule
\rowcolor{corvicsoft}\textbf{Corvic AI} & \textbf{1.16} & \textbf{1.04} & \textbf{2.19} & \textbf{20/20} & \textbf{4,300,000} & \textbf{2.47} & \textbf{14,800,000} \\
LoraDB & 0.59 & 0.64 & 2.60 & 18/20 & 5,000 & 247.00 & 1,500,000 \\
Ladybug (copy) & 1.65 & 1.56 & 4.08 & 20/20 & 476,000 & 0.96 & 2,400,000 \\
DuckPGQ & 1.37 & 1.56 & 4.18 & 20/20 & 3,500,000 & 3.10 & 16,700,000 \\
Memgraph & 0.69 & 0.96 & 4.77 & 20/20 & 87,000 & 0.90 & 974,000 \\
Neo4j & 2.35 & 2.64 & 6.92 & 20/20 & 53,000 & 7.23 & 335,000 \\
HugeGraph & 15.98 & 19.76 & 61.81 & 19/20 & 64,000 & 30.00 & DNF \\
\bottomrule
\end{tabular}

\vspace{4pt}
\footnotesize Note: batch update is throughput (rows/s) for a fixed-size mutation batch applied after initial load, distinct from the one-time bulk-ingest figure in the adjacent column; HugeGraph's cell is DNF (did not finish) under our harness at this scale.
\end{table}

\subsection{Disclosed exclusions and configuration notes}
\begin{itemize}[leftmargin=1.1em,itemsep=3pt,topsep=2pt]
\item \textbf{FalkorDB.} Evaluated; query geomean at 1.02M nodes was 1.81~ms with 18/20 shapes completing correctly (excluded from full-workload comparisons requiring 20/20 correctness; included in the point-update comparison, where it was measured at 0.85~ms).
\item \textbf{PuppyGraph.} Evaluated and excluded from all figures and tables in this paper; the system did not complete the full twenty-shape workload at the tested scales under our harness.
\item \textbf{Ladybug (attach vs.\ copy).} Two configurations were benchmarked. The \texttt{copy} configuration (reported in the main text and Table~\ref{tab:main}) achieved 793{,}462~rows/s ingest and 4.08~ms query geomean at 1.02M nodes. The \texttt{attach} configuration achieved 476{,}360~rows/s ingest and 3.97~ms query geomean at the same scale, with a faster point-update time (0.96~ms vs.\ 1.57~ms). Only one configuration is shown in the main-text charts to avoid representing one physical system as two independent data points.
\item \textbf{HugeGraph batch update.} Could not be measured to completion (DNF) at the full 1.02M-node scale under our harness and is omitted from the batch-update figure and the crossover analysis.
\item \textbf{Memgraph point-update caveat.} Memgraph's 0.90~ms point-update figure and 0.85~ms for FalkorDB are both faster than Corvic AI's 2.47~ms; this is the one metric in this study where Corvic AI does not lead among fully-completing engines, and is reported without adjustment.
\end{itemize}


\begin{thebibliography}{9}
\bibitem{google2012} Google. ``Introducing the Knowledge Graph: things, not strings.'' \textit{Google Blog}, 2012.
\bibitem{sqlpgq2024} ISO/IEC 39075:2024. \textit{Information technology --- Database languages --- SQL/PGQ (Property Graph Queries)}.
\bibitem{duckpgq2024} DuckPGQ Project. \textit{SQL/PGQ extension for relational query engines}. duckpgq.org, 2024.
\bibitem{ldbc2020} Angles, R. et al. ``The LDBC Social Network Benchmark.'' \textit{ACM SIGMOD Record}, 2020.
\bibitem{graphalytics2016} Iosup, A. et al. ``LDBC Graphalytics: A Benchmark for Large-Scale Graph Analysis.'' \textit{VLDB}, 2016.
\bibitem{corvicmem2026} Corvic AI. \textit{Memories: an ontology management system}. Internal technical documentation, 2026.
\bibitem{codd1970} E. F. Codd. ``A Relational Model of Data for Large Shared Data Banks.'' \textit{Communications of the ACM}, 13(6):377--387, 1970.
\bibitem{selinger1979} P. G. Selinger, M. M. Astrahan, D. D. Chamberlin, R. A. Lorie, T. G. Price. ``Access Path Selection in a Relational Database Management System.'' \textit{SIGMOD}, 1979.
\bibitem{stonebraker2005cstore} M. Stonebraker, D. J. Abadi, A. Batkin, et al. ``C-Store: A Column-oriented DBMS.'' \textit{VLDB}, 2005.
\bibitem{boncz2005x100} P. Boncz, M. Zukowski, N. Nes. ``MonetDB/X100: Hyper-Pipelining Query Execution.'' \textit{CIDR}, 2005.
\bibitem{abadi2008columnstore} D. J. Abadi, S. R. Madden, N. Hachem. ``Column-Stores vs. Row-Stores: How Different Are They Really?'' \textit{SIGMOD}, 2008.
\bibitem{francis2018cypher} N. Francis, A. Green, P. Guagliardo, et al. ``Cypher: An Evolving Query Language for Property Graphs.'' \textit{SIGMOD}, 2018.
\bibitem{rodriguez2015gremlin} M. A. Rodriguez. ``The Gremlin Graph Traversal Machine and Language.'' \textit{DBPL}, 2015.
\bibitem{malewicz2010pregel} G. Malewicz, M. H. Austern, A. J. C. Bik, et al. ``Pregel: A System for Large-Scale Graph Processing.'' \textit{SIGMOD}, 2010.
\bibitem{besta2023demystifying} M. Besta, R. Gerstenberger, E. Peter, et al. ``Demystifying Graph Databases: Analysis and Taxonomy of Data Organization, System Designs, and Graph Queries.'' \textit{ACM Computing Surveys}, 56(2), 2023.
\bibitem{sakr2021futurebiggraphs} S. Sakr, A. Bonifati, H. Voigt, A. Iosup, et al. ``The Future Is Big Graphs: A Community View on Graph Processing Systems.'' \textit{Communications of the ACM}, 64(9):62--71, 2021.
\bibitem{mcsherry2015cost} F. McSherry, M. Isard, D. G. Murray. ``Scalability! But at What COST?'' \textit{HotOS}, 2015.
\bibitem{fleming1986geomean} P. J. Fleming, J. J. Wallace. ``How Not to Lie with Statistics: The Correct Way to Summarize Benchmark Results.'' \textit{Communications of the ACM}, 29(3):218--221, 1986.
\bibitem{fagin2001topk} R. Fagin, A. Lotem, M. Naor. ``Optimal Aggregation Algorithms for Middleware.'' \textit{PODS}, 2001.
\bibitem{gruber1993ontology} T. R. Gruber. ``A Translation Approach to Portable Ontology Specifications.'' \textit{Knowledge Acquisition}, 5(2):199--220, 1993.
\bibitem{lewis2020rag} P. Lewis, E. Perez, A. Piktus, et al. ``Retrieval-Augmented Generation for Knowledge-Intensive NLP Tasks.'' \textit{NeurIPS}, 2020.
\bibitem{pan2024unifyingllmkg} S. Pan, L. Luo, Y. Wang, C. Chen, J. Wang, X. Wu. ``Unifying Large Language Models and Knowledge Graphs: A Roadmap.'' \textit{IEEE Transactions on Knowledge and Data Engineering}, 36(7):3580--3599, 2024.
\bibitem{edge2024graphrag} D. Edge, H. Trinh, N. Cheng, et al. ``From Local to Global: A Graph RAG Approach to Query-Focused Summarization.'' \textit{arXiv:2404.16130}, 2024.
\bibitem{angles2008survey} R. Angles, C. Gutierrez. ``Survey of Graph Database Models.'' \textit{ACM Computing Surveys}, 40(1), 2008.
\bibitem{wu2024stark} S. Wu, S. Zhao, M. Yasunaga, K. Huang, K. Cao, Q. Huang, V. N. Ioannidis, K. Subbian, J. Zou, J. Leskovec. ``STaRK: Benchmarking LLM Retrieval on Textual and Relational Knowledge Bases.'' \textit{NeurIPS Datasets and Benchmarks Track}, 2024.
\bibitem{chandak2023primekg} P. Chandak, K. Huang, M. Zitnik. ``Building a Knowledge Graph to Enable Precision Medicine.'' \textit{Scientific Data}, 10:67, 2023.
\end{thebibliography}
\end{document}